\documentclass[twocolumn,showpacs,aps,prl,superscriptaddress]{revtex4}

\usepackage{xspace}
\usepackage{graphicx}
\usepackage{dcolumn}
\usepackage{amsmath}
\usepackage{epsfig}

\input babarsym.tex

\newcommand{\eeeeg}   {\ensuremath{e^+e^- \to e^+e^-\gamma}\xspace}
\newcommand{\eett}   {\ensuremath{e^+e^- \to \tautau}\xspace}
\newcommand{\eeqq}       {\ensuremath{e^+e^- \to \qqbar}\xspace}

\newcommand{\tautoel}  {\ensuremath{\tau \to e \nunub}\xspace}
\newcommand{\tautoelg}  {\ensuremath{\tau \to e \nunub\gamma}\xspace}

\newcommand{\taumg}      {\ensuremath{\mtau^{\pm} \to \mmu^{\pm} \g}\xspace}
\newcommand{\taueg}      {\ensuremath{\mtau^{\pm} \to \electron^{\pm} \g}\xspace}
\newcommand{\muelg}      {\ensuremath{\mmu^{+} \to \electron^{+} \g}\xspace}

\newcommand{\taumgnch}      {\ensuremath{\mtau \to \mmu \g}\xspace}
\newcommand{\tauegnch}      {\ensuremath{\mtau \to \electron \g}\xspace}
\newcommand{\muelgnch}      {\ensuremath{\mmu \to \electron \g}\xspace}

\newcommand{\BRtaumg}    {\ensuremath{\BR(\taumg)}\xspace}
\newcommand{\BRtaueg}    {\ensuremath{\BR(\taueg)}\xspace}
\newcommand{\BRmuelg}    {\ensuremath{\BR(\muelg)}\xspace}

\newcommand{\roots}        {\ensuremath{\sqrt{s}}\xspace}

\newcommand{\ntaupair}         {\ensuremath{2.07\times 10^8}\xspace}

\newcommand{\lumion}             {\ensuremath{210.6 \invfb}\xspace}
\newcommand{\lumioff}             {\ensuremath{21.6 \invfb}\xspace}
\newcommand{\UpperLimitBarlow}        {\BRtaueg$<1.1\times10^{-7}$\xspace}
\newcommand{\UpperLimitExpected}        {\BRtaueg$<1.2\times10^{-7}$\xspace}
\newcommand{\Eeg}             {\ensuremath{E_{\electron\gamma}}\xspace}
\newcommand{\Mnu}             {\ensuremath{m_{\nu}^2}\xspace}
\newcommand{\ptmiss}            {\ensuremath{p^{T}_{\mathrm{miss}}}\xspace}
\newcommand{\logmisspt}         {\ensuremath{-\ln (2\times p^T_{miss}/\sqrt{s})}\xspace}
\newcommand{\EoverP}          {\ensuremath{E/p}\xspace}
\def\egam  {\ensuremath{\electron\gamma}\xspace}
\def\eff   {\ensuremath{\varepsilon}\xspace}
\def\DeltaEg {\ensuremath{\Delta E_\g}\xspace}

\def\kk       {\mbox{\tt KK2F}\xspace}
\def\tauola     {\mbox{\tt TAUOLA}\xspace}

\def\koralb     {\mbox{\tt KORALB}\xspace}

\newcommand{\gevccgevcc}{\ensuremath{{\mathrm{\,Ge\kern -0.1em V^2\!/}c^4}}\xspace}

\newcommand{\evcc}{\ensuremath{{\mathrm{\,e\kern -0.1em V\!/}c^2}}\xspace}
\newcommand{\CM} {\mbox{c.m.}\xspace}

\newcommand{\BABARPubYear}     {05}
\newcommand{\BABARPubNumber}  {040}

\newcommand{\SLACPubNumber} {11385}
\newcommand{\LANLNumber}  {0508012}

\def\figurebox#1#2#3{%
    \def\arg{#3}%
    \ifx\arg\empty
    {\hfill\vbox{\hsize#2\hrule\hbox to #2{\vrule\hfill\vbox to #1{\hsize#2\vfill}\vrule}\hrule}\hfill}%
    \else
    {\hfill\epsfbox{#3}\hfill}%
    \fi}

\begin{document}

\preprint{\babar-PUB-\BABARPubYear/\BABARPubNumber} 
\preprint{SLAC-PUB-\SLACPubNumber} 

\begin{flushleft}
\babar-PUB-\BABARPubYear/\BABARPubNumber\\
SLAC-PUB-\SLACPubNumber\\
hep-ex/\LANLNumber\\[10mm]    
\end{flushleft}

\title{
{\large \bf \boldmath
Search for Lepton Flavor Violation in the Decay \taueg} 
}

%
\author{B.~Aubert}
\author{R.~Barate}
\author{D.~Boutigny}
\author{F.~Couderc}
\author{Y.~Karyotakis}
\author{J.~P.~Lees}
\author{V.~Poireau}
\author{V.~Tisserand}
\author{A.~Zghiche}
\affiliation{Laboratoire de Physique des Particules, F-74941 Annecy-le-Vieux, France }
\author{E.~Grauges}
\affiliation{IFAE, Universitat Autonoma de Barcelona, E-08193 Bellaterra, Barcelona, Spain }
% ...
\collaboration{The \babar\ Collaboration}
\noaffiliation

\date{\today}

\begin{abstract}
A search for the non-conservation of lepton flavor 
in the decay \taueg has been performed with \ntaupair \eett events 
collected by the \babar\ detector at the PEP-II storage ring
at a center-of-mass energy near 10.58\gev.
We find no evidence for a signal and 
set an upper limit on the branching ratio of \BRtaueg\ $<1.1\times10^{-7}$ at 90\% confidence level.
\end{abstract}

\pacs{13.35.Dx, 11.30.Hv, 14.60.Fg}

\maketitle


Lepton flavor conservation differs from other
conservation laws in the Standard Model (SM)
because it is not associated with an underlying conserved current symmetry.
Consequently, new theories attempting to describe nature beyond the SM
often include lepton flavor violating processes
such as the neutrino-less decay of a $\mu$ or \mtau lepton, which have
long been identified as unambiguous signatures of new physics.
If no specific theoretical model is assumed, 
any or all of the \muelgnch, \taumgnch and \tauegnch decays can be expected to be observed,
and therefore independent searches for each of these modes are required.
Some theoretical models~\cite{Ma:2002pq,Ellis:2002fe} respecting the current limits 
on \BRmuelg~\cite{Brooks:1999pu} and \BRtaumg~\cite{Aubert:2005ye} 
in fact allow \taueg decays to occur up to the existing experimental bound~\cite{Hayasaka:2005xw}.

A significant improvement on this \taueg limit is presented here using 
data recorded by the \babar\ detector at the SLAC \pep2 asymmetric-energy \epem storage ring.
The data sample consists of an integrated luminosity of \L = \lumion\ recorded at a
center-of-mass (\CM) energy (\roots) of $\sqrt{s}=10.58\gev$, and \lumioff\ recorded at  $\sqrt{s}=10.54\gev$.
With an average cross section of $\sigma_{\eett}$ = (0.89$\pm$0.02) nb~\cite{kkxsec} as determined using the \kk Monte Carlo (MC) generator~\cite{kk}, 
this corresponds to a data sample of \ntaupair $\tau$-pair events.

The \babar\ detector is described in detail in Ref.~\cite{detector}.
Charged particles are reconstructed as tracks with 
a 5-layer silicon vertex tracker and a 40-layer drift chamber (DCH)
inside a 1.5 T solenoidal magnet.
An electromagnetic calorimeter (EMC) consisting of 6580 CsI(Tl) 
crystals is used to identify electrons and photons.
A ring-imaging Cherenkov detector (DIRC) is used to identify
charged hadrons and provides additional electron identification information.

The signature of the signal process is the presence of an isolated \egam 
pair having an invariant mass consistent with that of the \mtau (1.777\gevcc~\cite{bes}) and a total
energy (\Eeg) equal to \roots/2 in the \CM frame, 
along with other particles in the event with properties consistent with a SM \mtau decay.
Such events are simulated with higher-order radiative corrections using the \kk MC generator~\cite{kk} 
where one $\tau$ decays into \egam according to phase space~\cite{flatphasespace},
while the other $\tau$ decays according to measured branching ratios~\cite{Eidelman:2004wy} 
simulated with the \tauola MC generator~\cite{tauola,photos}.
The detector response is simulated with the \mbox{\tt GEANT4} package~\cite{geant}. 
The simulated events for signal as well as SM background
processes~\cite{kk,tauola,photos,Lange:2001uf,Sjostrand:1995iq,Jadach:1995nk}
are then reconstructed in the same manner as data.
The MC backgrounds are used to optimize the selection criteria and study systematic errors in the efficiency estimates,
but not for the estimation of the final background rate, which relies solely on data.
For the background from Bhabha events, we do not rely upon MC predictions
because the large Bhabha cross section makes generation of a sufficiently large MC sample impractical.

Events with zero total charge and with two or four well-reconstructed tracks inconsistent with coming from a photon conversion are selected.
The event is divided into hemispheres by the plane perpendicular
to the thrust axis. The thrust axis,
which characterizes the direction of maximum energy flow in the \CM frame of the event~\cite{thrust},
is calculated using all observed charged and neutral particles.

The signal-side hemisphere is required
to contain at least one \g\ with a \CM\ energy greater than 500\mev, and 
one track identified as an electron. The electron identification uses DCH, EMC and DIRC
information, including a requirement that the \EoverP ratio (the energy deposited in the EMC 
by the charged particle divided by its momentum as measured in the DCH) lies between 0.89 and 1.2.
The electron candidate is required to lie within the fiducial acceptance of the EMC
and to have a momentum greater than 500\mevc.
These criteria yield a $\pi$ mis-identification rate of less than 0.3\%. 
The efficiency for correctly identifying reconstructed tracks in
the fiducial volume as electrons in \taueg MC events is greater than 91\%.
For events with more than one signal-side $\gamma$ candidate,
we choose the $\gamma$ which gives the mass of the \egam system closest to the \mtau mass. 
This provides the correct pairing for $99.9\%$ of selected signal MC events.

\begin{figure}
 \resizebox{.96\columnwidth}{.27\textheight}{
\includegraphics{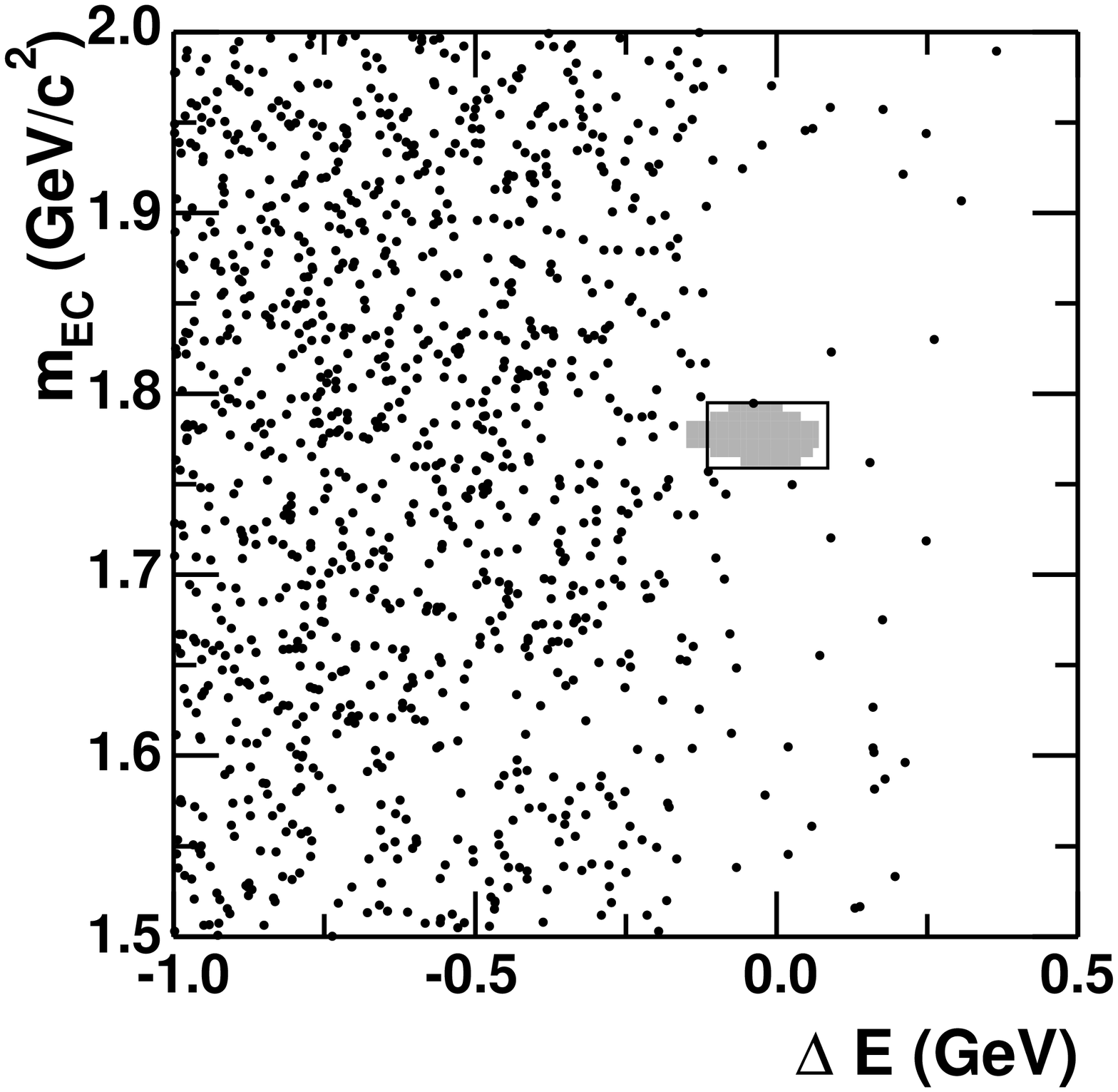}}
\caption{\mec vs. \DeltaE distribution of data (dots)
 and shaded region containing 50\% of the selected signal MC events inside the Grand Signal Box, as defined in the text.
 The boundary of the $\pm$2$\sigma$ signal box is also shown.}
\label{fig1}
\end{figure}

The resolution of the \egam mass is improved  by assigning 
the point of closest approach of the \electron track to the \epem collision axis
as the origin of the \g candidate and by using a kinematic fit with 
\Eeg\ constrained to \roots/2. The resulting energy-constrained mass (\mec) and
$\DeltaE =\Eeg -\roots/2$ are independent variables 
apart from small correlations arising from initial and final state radiation.
The mean and standard deviation of the \mec and \DeltaE distributions for reconstructed 
MC signal events are: $\langle \mec \rangle$ = 1777\mevcc, $\sigma(\mec)$ = 9\mevcc, 
$\langle \DeltaE \rangle$ = $-$15\mev, $\sigma(\DeltaE)$ = 51\mev, 
where the shift in $\langle \DeltaE \rangle$ comes from photon energy reconstruction effects.
To minimize possible biases, we perform a blind analysis by excluding all events in the data within a $\pm$3$\sigma$ rectangular 
box centered on $\langle \mec \rangle$ and $\langle \DeltaE \rangle$
until all optimization and systematic studies of the selection criteria have been completed.
We optimize the selection to obtain the smallest expected upper limit 
in a background-only hypothesis for observing events inside a $\pm2\sigma$ rectangular box signal box defined by:
$|\DeltaE - \langle \DeltaE \rangle| < 2 \sigma(\DeltaE)$ and $|\mec - m_{\tau}| < 2 \sigma(\mec)$,
as shown in Figure~\ref{fig1}.

The dominant backgrounds arise from Bhabha and \eett (with a \tautoel\ decay) processes
with an energetic $\gamma$ from initial or final state radiation or from \tautoelg decays.
Backgrounds arising from radiation are reduced by requiring that the total \CM\ energy
of all non-signal $\gamma$ candidates in the signal-side hemisphere be less than 200\mev.
To suppress non-$\tau$ backgrounds with significant radiation along the beam directions,
the polar angle $(\theta_{miss})$ of the missing momentum associated with the neutrino(s) in the event
is required to lie within the detector acceptance $(-0.76 < \cos\theta_{miss} < 0.92)$.

The tag-side hemisphere, defined to be that opposite to the signal-side hemisphere,
 is expected to contain a SM \mtau decay  characterized by the presence of 
one or three charged particles and missing momentum due to unobserved neutrino(s).
Taking the direction of the tag-side $\tau$ to be opposite the signal $\egam$ candidate,
we use all tracks and $\gamma$ candidates in the tag-side hemisphere to calculate the 
invariant mass squared of the tag-side missing momentum (\Mnu), which peaks around zero for the signal.
To reduce backgrounds from radiative \eett processes, we require \Mnu $>$ $-0.25$ \gevccgevcc.

The component of the missing momentum of the event transverse to the collision axis scaled to the beam energy
$(2 \times p^{T}_{miss}/\sqrt{s})$ is expected to be large for signal and \eett events, 
but small for Bhabha and 2-photon events.
We exploit an observed correlation between \Mnu and $(2 \times p^{T}_{miss}/\sqrt{s})$ 
in the non-$\tau$ backgrounds to significantly suppress them.
We require the following: $(\Mnu/1.8\gevccgevcc)\logmisspt/2.0 < 1$,
the highest \CM\ momentum track on the tag-side hemisphere 
to be inconsistent with being an electron, including requirements
that \EoverP be less than 0.5 and that the momentum be greater than 500\mevc,
and the tag-side hemisphere to have a total \CM\ momentum of all
 charged and neutral particles less than 4.75\gevc.

Backgrounds from \eeqq processes are further reduced by requiring the
 total invariant mass of particles in the tag-side hemisphere to be
less than 1.8\gevcc.

After this selection, 8.9\% of the total generated MC signal events survive within 
a Grand Signal Box (GSB) region defined as follows: 
\mec $\in$ $[1.5, 2.0]$ \gevcc, \DeltaE $\in$ $[-1.0, 0.5]$ \gev.
The data distribution of \mec and \DeltaE inside the GSB is plotted as dots in Figure~\ref{fig1},
along with a shaded region containing 50\% of the selected signal MC events shown for illustrative purposes.
The GSB excluding the $\pm3\sigma$ blind region contains 1110 data events,
while the luminosity-normalized sum of the non-Bhabha MC backgrounds yield 1045 events.
Of these MC events, 99.8\% are \eett events, 99.9\% of which have \tautoel decays on the signal-side.

The (5.9$\pm$3.7)\% difference between the number of data and $\tau$-pair dominated MC events
indicates that the Bhabha background level in the GSB is low. 
However, in the more restrictive  $|\DeltaE - \langle \DeltaE \rangle| < 2 \sigma(\DeltaE)$ region, 
the Bhabha background is expected to contribute a substantially higher background fraction because of the
greater likelihood of a Bhabha than a $\tau$-pair event to have a hemisphere containing the full beam energy.
This residual Bhabha contamination is studied using data distributions of the deviation (\DeltaEg)
of the measured photon \CM\ energy from the corresponding prediction assuming a fully contained 
\eeeeg event.
The predicted photon energy is obtained from the beam energy and kinematic information
from all particles in the event except the measured photon energy.
We observe that the excess of data over non-Bhabha MC events is clustered at low \DeltaEg,
where the Bhabha events are expected to appear.
As we progressively loosen the electron veto on the tag-side track, the excess in the number of data events
over the non-Bhabha MC background grows in the region with small \DeltaEg, 
providing further confirmation that the Bhabha background is well understood.

We cross check the Bhabha contamination 
in the  $|\DeltaE - \langle \DeltaE \rangle| < 2 \sigma(\DeltaE)$ region
from a data sample without a tag-side electron veto,
by removing the \EoverP requirement on the tag-side.
To estimate the Bhabha contamination surviving our final event selection, 
which includes a cut of tag-side $\EoverP<0.5$,
we use the data in the adjacent Bhabha-dominated \EoverP region, $0.5<\EoverP<1.2$.
We extrapolate the rate from the $0.5<\EoverP<1.2$ region to the $\EoverP<0.5$ region, 
using a high statistics and high purity Bhabha control sample obtained by reversing the requirement on
$(\Mnu/1.8\gevccgevcc)\logmisspt/2.0$ given above.
We estimate the residual Bhabha contamination in our final selection 
by multiplying the number of events in the $0.5<\EoverP<1.2$ region of the no tag-side electron veto sample 
by the ratio of the number of events in the Bhabha control sample in the $\EoverP<0.5$ region
to that in the $0.5<\EoverP<1.2$ region.
This method gives an estimate of 10.3$\pm$1.1 Bhabha events
inside the $\pm$2$\sigma$(\DeltaE) band once the tag-side electron veto is applied.

\begin{figure}
 \resizebox{.96\columnwidth}{.27\textheight}{
\includegraphics{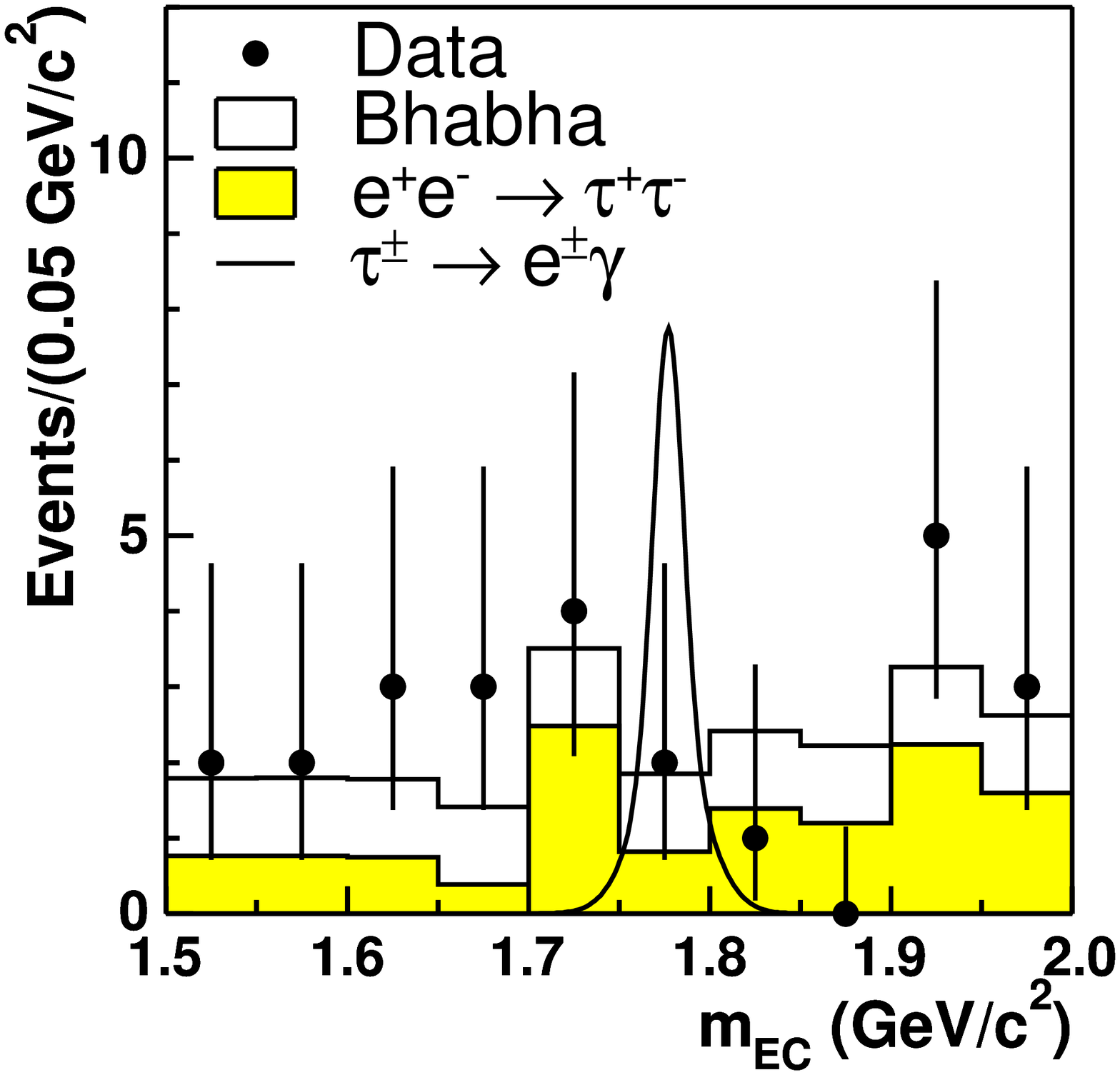}}
\caption{\mec distribution of data (dots),
 the expected backgrounds (histograms) and
 MC signal (curve with arbitrary normalization) 
 for $|\DeltaE - \langle \DeltaE \rangle| < 2 \sigma$(\DeltaE).}
\label{fig2}
\end{figure}

In this band, we expect 12.9$\pm$2.5 events from the non-Bhabha MC backgrounds,
thus obtaining a total background estimate of 23.2$\pm$2.7 events.
This compares well with the 25 events observed inside the $\pm$2$\sigma$(\DeltaE) band in the data.
We also find good agreement between the observed and expected number of events 
separately for the sub-samples with one and three tracks on the tag-side.

For the final background estimate we use the \mec distribution of data events
inside the $\pm$2$\sigma$(\DeltaE) band,
as shown in Figure~\ref{fig2}
along with the signal shape included for illustrative purposes.
The backgrounds from data inside the $\pm$2$\sigma$(\DeltaE) band 
with $|\mec - m_{\tau}| > 3 \sigma(\mec)$
are fitted to different orders of polynomials in \mec using a maximum likelihood approach.
A fit with a constant probability density function (PDF) yields a total $\chi^2$ of 4.7 for
the 10 bins shown in Figure~\ref{fig2},
and predicts 1.9$\pm$0.4 events inside the final $\pm$2$\sigma$(\mec) signal region.
Equally acceptable goodness of fit is obtained with higher-order polynomials.
However, the coefficients of the higher order terms are statistically compatible with zero.
The background predictions from these PDFs agree with the prediction from the constant PDF to within $\pm$0.3 events.
As these deviations are smaller than the statistical error on the prediction from the constant PDF,
we conclude that the data \mec distribution is consistent with being uniform.

A cross check using non-Bhabha MC background contributions combined with residual Bhabha contamination estimates
obtained from the data is also found to be reasonably uniform in \mec (Figure~\ref{fig2})
and predicts 1.7$\pm$0.2 events  inside the $\pm$2$\sigma$(\mec) signal box.

The (5.9$\pm$3.7)\% difference  between data and $\tau$-pair MC predictions 
also provides a measure of our ability to model the signal-like events in the GSB,
since these data events have very similar characteristics to the signal,
both in terms of the trigger response of the experiment as well as for the distributions of all the selection variables
apart from \mec and \DeltaE. 
The systematic error due to a particular cut is taken as the product of the marginal efficiency of the cut and the
relative discrepancy between data and MC in the GSB after all other cuts have been applied.
The contributions from all the different cuts added in quadrature yield a 2.3\% relative systematic error,
the only appreciable effect being associated with the requirements on \Mnu and \ptmiss.
This approach yields a more conservative estimate of the systematic uncertainty on the signal efficiency
than the more traditional approach derived from considering the difference between the data and MC prediction
for each selection variable, which gives a total estimate of 2.0\% relative contribution from all the cuts.

The relative systematic uncertainties on the trigger efficiency,
tracking and photon reconstruction efficiencies, and particle identification
are estimated to be 1.4\%, 1.3\%, 1.8\% and 1.3\%, respectively.
The requirement that the events fall within the $\pm$2$\sigma$ signal box in \mec and \DeltaE
contributes a 4.4\% systematic error associated with the scale and resolution uncertainties of
these variables and a small contribution from the beam energy uncertainty.
As we use 1.3 million MC signal events, the contribution to the
uncertainty arising from signal MC statistics is negligible.
Adding the contributions of the individual terms in quadrature with
an additional 2.3\% normalization error on the product $\L\sigma_{\tau\tau}$
gives a 6.2\% total relative systematic uncertainty on $\L\sigma_{\tau\tau}$\eff in
the signal box, where the efficiency is \eff = (4.7$\pm$0.3)\%.
We note that our final limit on the branching ratio is insensitive to the
systematic uncertainty as long as this uncertainty is below 10\%.

We find one event in the signal box for an expected background of 1.9$\pm$0.4 events.
Because of the low background levels, we do not fit for a signal in the \mec distribution
as is done in our recent search for \taumg\ ~\cite{Aubert:2005ye}. Rather, we set an upper limit
employing the same technique used in our  search for $\tau^{\pm}\ra\ell^{\pm}\ell^+\ell^-$~\cite{Aubert:2003pc}
where the background levels were also small.
A 90\% C.L. upper limit on the branching ratio is calculated according to 
$\BR^{90}_{UL}=N^{90}_{UL}/(2\eff\L\sigma_{\tau\tau})$, 
where $N^{90}_{UL}$ is the 90\% C.L. upper limit with one event observed
when 1.9$\pm$0.4 events are expected. 
The  limit  is calculated including all uncertainties
 using the technique of Cousins and Highland~\cite{Cousins:1992qz} 
following the implementation of Barlow~\cite{Barlow:2002bk}. 
At 90\% C.L. this procedure gives an upper limit of \UpperLimitBarlow~\cite{Sensitivity}.
This represents a more than three-fold reduction in the upper limit as reported in~\cite{Hayasaka:2005xw}.

\begin{figure}
 \resizebox{.96\columnwidth}{.2\textheight}{\includegraphics{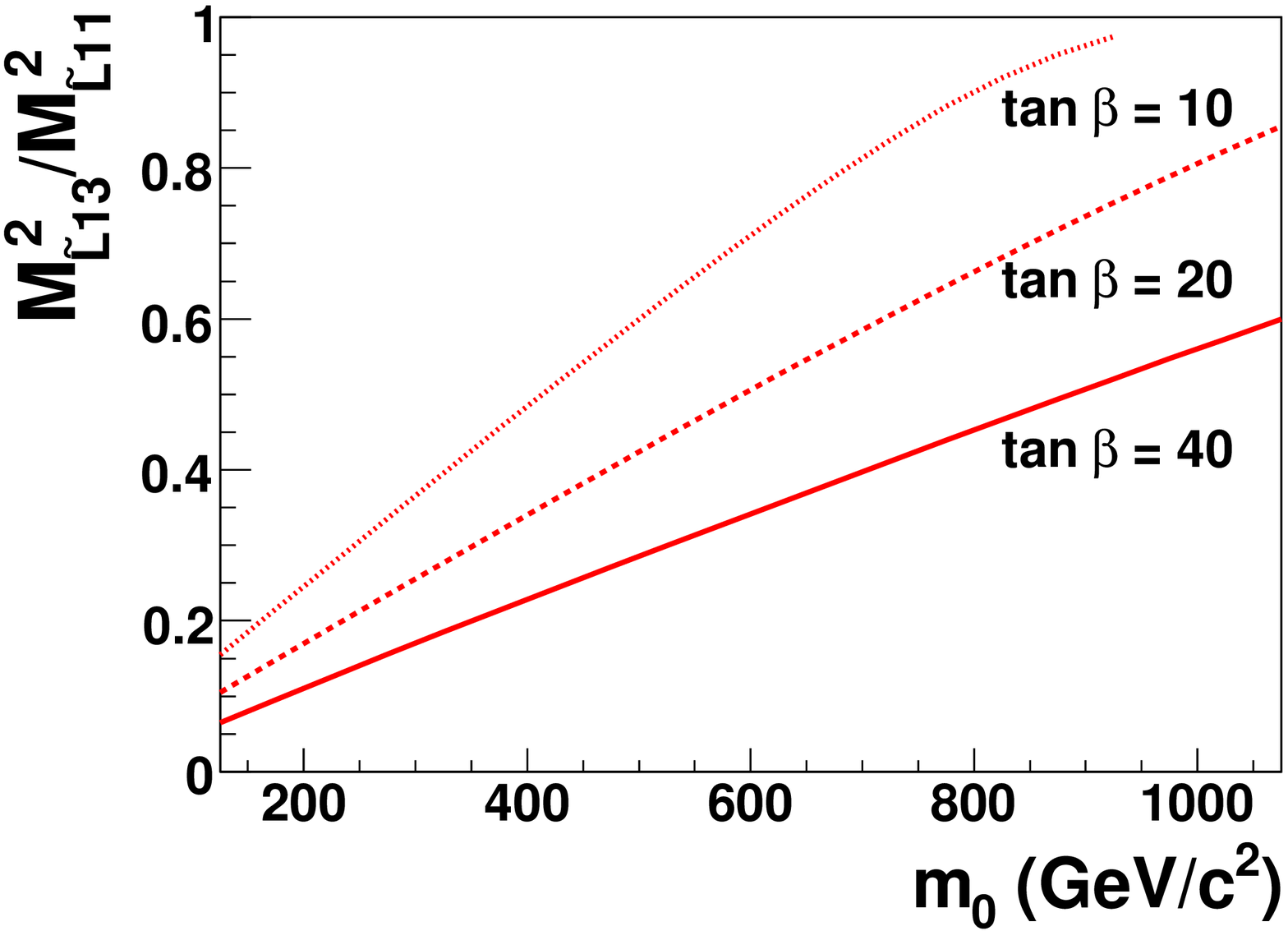}}
\caption{90\% C.L. upper limits on $M^2_{\tilde{L}13}/M^2_{\tilde{L}11}$ 
for \BRtaueg $<$ 1.1 $\times$ 10$^{-7}$ with $\tan\beta$ = 10, 20 and 40.}
\label{fig3}
\end{figure}

As an example of how this result constrains theories beyond the SM,
we set bounds on the ratio of the first and the third generation element 
to the first generation diagonal element ($M^2_{\tilde{L}13}/M^2_{\tilde{L}11}$) 
of the left-handed slepton mass matrix based on predictions from a minimal supergravity model~\cite{Brignole:2004ah,Porod:2003um}.
Figure~\ref{fig3} shows the upper limits on $M^2_{\tilde{L}13}/M^2_{\tilde{L}11}$
 as a function of the ratio of the Higgs vacuum expectation values 
($\tan\beta$) and the universal scalar mass ($m_0$), which, for simplicity, is set equal to the universal gaugino mass.

We are grateful for the excellent luminosity and machine conditions
provided by our \pep2\ colleagues, 
and for the substantial dedicated effort from
the computing organizations that support \babar.
The collaborating institutions wish to thank 
SLAC for its support and kind hospitality. 
This work is supported by
DOE
and NSF (USA),
NSERC (Canada),
IHEP (China),
CEA and
CNRS-IN2P3
(France),
BMBF and DFG
(Germany),
INFN (Italy),
FOM (The Netherlands),
NFR (Norway),
MIST (Russia), and
PPARC (United Kingdom). 
Individuals have received support from the 
A.~P.~Sloan Foundation, 
Research Corporation,
and Alexander von Humboldt Foundation.


\begin{thebibliography}{99}
\bibitem{flatphasespace}
Our limit is insensitive to the phase space assumption
as cross-checked with the two extreme cases of 
a V $-$ A and a V $+$ A form of interaction for the signal MC.
\end{thebibliography}
\end{document}